\documentstyle[aps,preprint]{revtex}
\tightenlines

\begin{document}
\draft
\title{ Landau Transport equations in slave-boson mean-field theory of $t-J$ model }
\author{Tai-Kai Ng}
\address{
Department of Physics,
Hong Kong University of Science and Technology,\\
Clear Water Bay Road,
Kowloon, Hong Kong
}
\date{ \today }
\maketitle
\begin{abstract}
  In this paper we generalize slave-boson mean-field theory for $t-J$ model to the time-dependent regime, and derive transport equations for $t-J$ model, both in the normal and superconducting states. By eliminating the boson and constraint fields exactly in the equations of motion we obtain a set of transport equations for fermions which have the same form as Landau transport equations for normal Fermi liquid and Fermi liquid superconductor, respectively with all Landau parameters explicity given. Our theory can be viewed as a refined version of U(1) Gauge theory where all lattice effects are retained and strong correlation effects are reflected as strong Fermi-liquid interactions in the transport equation. Some experimental consequences are discussed.

\end{abstract}


\narrowtext

\section{introduction}
  With accumulated experimental evidences it is now generally believed that high-$T_c$ superconductors are BCS-like d-wave superconductors close to Mott insulator. It is also believed that strong electron correlation does not destroy Fermi-liquid behaviour at low temperature, and the ground state of the superconducting states remains Fermi-liquid-(superconductor)\cite{e1,e2,e3} like. Theoretically strong electron correlation is often handled by introducing slave-particles, together with a constraint that the sum of electron plus slave particle is equal to one on every lattice site in the system. The method allows for approximate mean-field solution which gives qualitatively reasonable results in many cases\cite{sf,sb}. 

  Within the slave-particle approach a simple theory that describes a strongly-correlated d-wave superconductor is the $t-J$ model in the slave-boson formulation\cite{sbtj}. The slave-boson mean-field theory (SBMFT) of the model produces a phase diagram which agrees qualitatively with experiments\cite{sbtj}. To avoid complicated mathematics, fluctuations upon mean-field theory are often studied in the continuum limit in the language of gauge theories\cite{le1,le2,le3}. Whereas the low temperature phases of gauge theory approaches are Fermi-liquid like, non-Fermi liquid behaviour arises rather naturally in the formulation at high temperature, and can explain qualitatively a number of properties of cuprates\cite{le1,le2}. However, the gauge theory approach disagrees with experiments in some important details\cite{le3,mgil} and it is generally believed that the theory is not a realistic representation of high-$T_c$ cuprates in its present form.

 In this paper we perform a {\em tour de force} analysis of Gaussian fluctuations above SBMFT of $t-J$ model without introducing continuum approximation. We extend SBMFT for the $t-J$ model to the time-dependent regime and study Gaussian fluctuations in the model in form of transport equations for both normal and superconducting states. We shall restrict ourselves to the zero-temperature, bose-condensed phase in this paper where the mean field state is Fermi-liquid like\cite{le1,le2,le3}. By eliminating the boson and constraint fields exactly in the equations of motion we obtain a set of transport equations for fermions of the same form as transport equations for normal Fermi liquids and Fermi liquid superconductors\cite{np,legett}, respectively. Effective $\vec{q}$-dependent Landau interaction can be extracted directly from our transport equations. When computing physical response functions our approach gives more refined results compared with U(1) gauge theory\cite{le1,le2,le3} and offers a precise Fermi liquid interpretation to the results. Moreover, with the Fermi liquid representation our theory can be extended beyond Gaussian theory more readily. We shall show how strong Fermi liquid interactions which become singular in the Mott Insulator limit $x\rightarrow0$ ($x$= hole concentration) arise naturally as a result of strong correlation in the model and shall discuss some of their physical consequences. Unfortunately we discover that the major drawback of gauge theory is not remedied by this more complete form of Gaussian theory.

  We begin in section II with a brief revision of slave-boson mean-field theory for the $t-J$ model. In section III we generalize SBMFT to the time-dependent regime and derive transport equations in the linear response regime. By eliminating the boson (hole) and constraint fields exactly we obtain a set of transport equations in terms of fermions only. The equations have the same form as corresponding transport equations for Fermi liquids in the $\vec{q}\rightarrow0$ limit and we can extract $\vec{q}$-dependent Landau interactions from the transport equation explicitly.

  The solutions of the transport equations are discussed in section IV. With a simplified form of Landau interaction we compute the density and current response functions and show that our refinements does not modify qualitatively the results obtained from U(1) gauge theory (See also Appendix A for a direct derivation of gauge theory results). The important effects of screening and current renormalization are discussed. The origin of the small current carried by quasi-particles $j_p\sim x$\cite{le3,mi} in gauge theory is interpreted in Landau Fermi liquid theory language. Our results are summarized in section V where we comment on the lessons we learn and limitations of the theory.

\section{slave-boson mean-field theory for $t-J$ model}   
  We consider a generalized $t-J$ model on a square lattice\cite{sbtj} that includes Coulomb interaction between charges\cite{tjdlee}. In slave-boson representation the Hamiltonian is 
\begin{eqnarray}
\label{ham}
H & = & -t\sum_{<i,j>\sigma}\left(b_ib^+_jc^+_{i\sigma}c_{j\sigma}+H.C.\right)-
\mu\sum_{i\sigma}c^+_{i\sigma}c_{i\sigma}+J\sum_{<i,j>}\vec{S}_i.\vec{S}_j
+{1\over2}\sum_{i,j}V(\vec{r}_i-\vec{r}_j)b^+_ib_ib^+_jb_j  \\  \nonumber
& & +\sum_{i}\lambda_i\left(b^+_ib_i+\sum_{\sigma}c^+_{i\sigma}c_{i\sigma}-1\right),
\end{eqnarray}
where $c^+_{i\sigma}(c_{i\sigma})$ and $b^+_i(b_i)$ are the fermion (spin) and boson (hole) creation (annihilation) operators at site $i$, respectively. The physical electron operator is $f_{i\sigma}=c_{i\sigma}b^+_i$. $\vec{S}_i=\psi_i^+\vec{\sigma}\psi_i$, where $\vec{\sigma}$ are Pauli matrices and $\psi_i=\pmatrix{c_{i\uparrow} \cr c_{i\downarrow}}$. The first term in the Hamiltonian represents electron hopping where $<i,j>$ denotes nearest neighbor pairs on a square lattice. $\mu$ is the chemical potential that fix the average density of electrons. The third and forth terms represent Heisenberg exchange interaction between electron spins and repulsive Coulomb interaction between charges, respectively. The constraint that the number of electrons $+$ number of empty site $= 1$ on every site is introduced in the last term {\em via} the Lagrange multiplier field $\lambda_i$ in the Hamiltonian. 

   In SBMFT the various terms in the Hamiltonian are decoupled as follows\cite{sbtj}:
\begin{mathletters}
\label{decoupling}
\begin{equation}
\label{det}
b_ib^+_jc^+_{i\sigma}c_{j\sigma}\rightarrow<b_i><b^+_j>c^+_{i\sigma}c_{j\sigma}+b_ib^+_j<c^+_{i\sigma}c_{j\sigma}>-<b_i><b^+_j><c^+_{i\sigma}c_{j\sigma}>, 
\end{equation}
where $<b_i>=<b^+_i>=\bar{b}=\sqrt{x}$ at equilibrium, where $x=$ concentration of holes.
\begin{eqnarray}
\label{dej}
\vec{S}_i.\vec{S}_j & \rightarrow & -{3\over8}\left[<\Delta^+_{ij}>\Delta_{ij}
+\Delta^+_{ij}<\Delta_{ij}>-<\Delta^+_{ij}><\Delta_{ij}>\right.  \\  \nonumber
& & +\left.<\chi^+_{ij}>\chi_{ij}+\chi^+_{ij}<\chi_{ij}>-<\chi^+_{ij}><\chi_{ij}>\right],
\end{eqnarray}
where $<..>$ denotes expectation values. $\Delta_{ij}=c_{i\uparrow}c_{j\downarrow}-c_{i\downarrow}c_{j\uparrow}$ and $\chi_{ij}=\sum_{\sigma}c^+_{i\sigma}c_{j\sigma}$. The $\Delta_{ij}$ term represents decoupling of the Heisenberg interaction in Cooper channel and is responsible for superconductivity, whereas the $\chi_{ij}$ terms represent decoupling in "Fock" channel, and gives rise to self-energy corrections to the fermions. The boson (density-density) interaction term
is decoupled as 
\begin{equation}
\label{dec}
b^+_ib_ib^+_jb_j\rightarrow<b>^2(b^+_i+b_i)(b^+_j+b_j), 
\end{equation}
which is the usual Bogoliubov decoupling\cite{mahan}. The Lagrange multiplier field is replaced by a constant number at equilibrium,
\begin{equation}
\label{dec}
\lambda_i\rightarrow<\lambda_i>=\bar{\lambda}.  
\end{equation}
\end{mathletters}

  The resulting mean-field Hamiltonian describes two separate species of particles. The mean-field Hamiltonian for fermions (spins) describes a BCS-superconductor with 
\begin{equation}
\label{hmfs}
H_{MF}^f=\sum_{\vec{k}\sigma}\xi_{\vec{k}}c^+_{\vec{k}\sigma}
c_{\vec{k}\sigma}+\sum_{\vec{k}}\left[\bar{\Delta}^*(\vec{k})(c_{\vec{k}\uparrow}
c_{-\vec{k}\downarrow}-c_{\vec{k}\downarrow}c_{-\vec{k}\uparrow})+H.C.\right],
\end{equation}
where $\xi_{\vec{k}}=-(t\bar{b}^2+{3J\over8}\bar{\chi})\gamma(\vec{k})
+\bar{\lambda}-\mu$, $\bar{\chi}=<\chi_{ij}>$ where $j=i+\hat{\mu} (\mu=x,y)$, and 
$\gamma(\vec{k})=2(cos(k_x)+cos(k_y))$. $\bar{\Delta}(\vec{k})={3J\over4}\bar{\Delta}(cos(k_x)-cos(k_y))$ represents a fermion pairing field with $d_{x^2-y^2}$ symmetry, where $\bar{\Delta}=<\Delta_{i,i+\hat{x}}>$. ${3J\over8}\bar{\chi}\gamma(\vec{k})$ is the "Fock" self-energy from the Heisenberg interaction, whereas $\bar{\Delta}(\vec{k})$ comes from the "Cooper" channel decoupling. The mean-field dispersion for the fermion quasi-particles is $E_f(\vec{k})=\pm\sqrt{\xi_{\vec{k}}^2+|\bar{\Delta}(\vec{k})|^2}$\cite{sbtj}. Notice that the effective hopping from $t$-term is renormalized by a factor $\bar{b}^2\sim x$ in mean-field theory and the fermion dispersion is dominated by the "Fock"
self=energy.

  The boson (hole) mean-field Hamiltonian is
\begin{equation}
\label{hmfh}
H_{MF}^h=\sum_{\vec{q}}\epsilon(\vec{q})b^+_{\vec{q}}b_{\vec{q}}
+{\bar{b}^2\over2}\sum_{\vec{q}}V(\vec{q})(b^+_{\vec{q}}+b_{-\vec{q}})(b_{\vec{q}}
+b^+_{-\vec{q}}),
\end{equation}
where $V(\vec{q})$ is the Fourier transform of $V(\vec{r})$ and $\epsilon(\vec{q})=-t\bar{\chi}\gamma(\vec{q})+\bar{\lambda}$. The bosons are bose-condensed in the ground state implying that $\bar{\lambda}=t\bar{\chi}\gamma(0,0)$. The mean-field dispersion for the density excitation is $E_h(\vec{q})=\sqrt{\epsilon(\vec{q})^2+2x\epsilon(\vec{q})V(\vec{q})}$. For short-ranged repulsive interaction, $E_h(\vec{q})\sim\sqrt{x}q$ at small $q$, whereas for (2D) Coulomb interaction, $E_h(\vec{q})\sim\sqrt{xq}$ (acoustic plasmon). The mean-field Hamiltonians together describe a normal Fermi liquid state when $\bar{\Delta}=0, <b>\neq0, \bar{\chi}\neq0$, and d-wave superconducting state when $\bar{\Delta}\neq0, <b>\neq0, \bar{\chi}\neq0$\cite{sbtj}. 

  We note that we have followed the so-called U(1) formulation of the $t-J$ model in this paper. An alternative formulation is the $SU(2)$ formulation\cite{le2}.  Mathematically the main difference between the two approaches is that the constraint of no-double occupancy is written in an SU(2) symmetric way in SU(2) theory, which introduces additional constraints $c_{i\sigma}c_{i-\sigma}=0$  and its Hermitian conjugate\cite{le2}. These additional constraints are automatically satisfied if the constraint $b^+_ib_i+\sum_{\sigma}c^+_{i\sigma}c_{i\sigma}=1$ is imposed exactly. In mean-field theories where constraints are only satisfied only on average, the introduction of SU(2) constraints results in more complicated transport equations which are probably more reliable. In the low-temperature superconducting states the SU(2) formulation predicts new collective modes\cite{ln} near momentum $\vec{q}\sim(\pi,\pi)$ that are not treated correctly in U(1) theory. However, the two approaches give qualitatively similiar results at $\vec{q}\sim(0,0)$\cite{ln}, which is the regime we shall concentrate on in this paper. We follow the U(1) formulation in this paper for simplicity.  

\section{time-dependent mean-field theory and transport equations for $t-J$ model}
   We now extend SBMFT to the time-dependent regime. The approach can be understood by considering the Heisenberg equations of motion (at imaginary time) ${\partial\over\partial\tau}\hat{O}=[H,\hat{O}]$ for operators $\hat{O}_2$'s that are quadratic in the fermion creation/annihilation operators. Because of the presence of four-operator terms in the Hamiltonian, the equations of motion of $\hat{O}_2$ will generate terms $\hat{O}_4$ which involve four-operator terms. The idea of (time-dependent) mean-field theory is to decouple the expectation value of the four-operator terms in terms of products of two-operator teams in the equation of motion, i.e. $<\hat{O}_4>\sim<\hat{O}_2^{(1)}><\hat{O}_2^{(2)}>$, thus arriving at self-consistent equations for expectation values of two-operator terms $<\hat{O}_2^{(i)}>$\cite{np}. The decoupling scheme for SBMFT at equilibrium states will be used in the Heisenberg equation of motion.

  We first consider equation of motion for the operator $\chi_{ij}$ defined in last section in the presence of external vector EM field $A^{ext}_{ij}(\tau)$ and scalar field $\Phi^{ext}_i(\tau)$. Following the above mentioned procedures, we obtain after some algebra the mean-field equation of motion,
\begin{mathletters}
\label{eqmotion}
\begin{eqnarray}
\label{em1}
{\partial\over\partial\tau}<\chi_{ij}(\tau)> & = & -{3J\over8}\sum_{j'=i+\hat{\mu}}
\left(<\Delta^+_{ij'}(\tau)><\Delta_{j'j}(\tau)>+<\chi_{ij'}(\tau)><\chi_{j'j}(\tau)>\right)  \\  \nonumber
& & +{3J\over8}\sum_{i'=j+\hat{\mu}}\left(<\Delta^+_{ii'}(\tau)><\Delta_{i'j}(\tau)>+<\chi_{ii'}(\tau)><\chi_{i'j}(\tau)>\right)  \\  \nonumber
& & -t\sum_{j'=i+\hat{\mu}}e^{iA^{ext}_{j'i}(\tau)}<b_{j'}(\tau)><b^+_i(\tau)><\chi_{j'j}(\tau)>  \\  \nonumber
& & +t\sum_{i'=j+\hat{\mu}}e^{iA^{ext}_{ji'}(\tau)}<b_j(\tau)><b^+_{i'}(\tau)><\chi_{ii'}(\tau)>  \\  \nonumber
& & +(<\lambda_i(\tau)>-<\lambda_j(\tau)>)<\chi_{ij}(\tau)>,
\end{eqnarray}
where various terms have same meaning as in last section except the introduction of time-dependence. The vector field $A^{ext}_{ij}(\tau)$ couples to the electrons through the hopping-term\cite{le1}. Similarly we obtain for the $\Delta_{ij}$ and $b_i$ operators,
\begin{eqnarray}
\label{em2}
{\partial\over\partial\tau}<\Delta_{ij}(\tau)> & = & -{3J\over8}\sum_{j'=i+\hat{\mu}}\left(<\Delta_{ij'}(\tau)><\chi_{j'j}(\tau)>-<\chi_{j'i}(\tau)><\Delta_{j'j}(\tau)>\right)  \\  \nonumber
& & +{3J\over8}\sum_{i'=j+\hat{\mu}}\left(<\Delta_{ji'}(\tau)><\chi^+_{ii'}(\tau)>+<\chi_{i'j}(\tau)><\Delta_{ii'}(\tau)>\right)  \\  \nonumber
& & +t\sum_{j'=i+\hat{\mu}}e^{iA^{ext}_{ij'}(\tau)}<b_i(\tau)><b^+_{j'}(\tau)><\Delta_{j'j}(\tau)>  \\  \nonumber
& & +t\sum_{i'=j+\hat{\mu}}e^{iA^{ext}_{ji'}(\tau)}<b_j(\tau)><b^+_{i'}(\tau)><\Delta_{ii'}(\tau)>  \\  \nonumber
& & -(<\lambda_i(\tau)>+<\lambda_j(\tau)>-2\mu)<\Delta_{ij}(\tau)>,
\end{eqnarray}
and 
\begin{equation}
\label{em3}
{\partial\over\partial\tau}<b_i(\tau)>=t\sum_{j'=i+\hat{\mu}}e^{iA^{ext}_{j'i}(\tau)}<b_{j'}(\tau)><\chi_{j'i}(\tau)>-(<\lambda_i(\tau)>+\Phi^{ext}_i(\tau))<b_i(\tau)>,
\end{equation}
\end{mathletters}
where we have coupled the external scalar potential to the bosons\cite{le1}. Similar equations of motion are also obtained for $\Delta^+_{ij}(\tau)$ and $b^+_i(\tau)$ fields. Notice that there is no equation of motion for the $<\lambda_i(\tau)>$ field. Instead, it should be chosen such that the average constraint $<b^+_i(\tau)><b_i(\tau)>+\sum_{\sigma}<c_{i\sigma}^+(\tau)c_{i\sigma}(\tau)>=1$ is satisfied at all sites $i$'s at all time $\tau$.

  To derive transport equations we go to wave-vector space and linearize the equation of motion. First we introduce our notations for the Fourier transformed fields. We write
$<\chi_{ij}(\tau)>=\bar{\chi}_{ij}+\delta\chi_{ij}(\tau)$, where $\bar{\chi}_{ij}$ is the equilibrium value of $<\chi_{ij}>$ and $\delta\chi_{ij}(\tau)$ is the time-dependent fluctuation. The Fourier transform field is defined by
\begin{mathletters}
\label{ft}
\begin{equation}
\label{fourier1}
FT[\chi_{ij}]=\sum_{i,j}e^{-i(\vec{k}+\vec{q}/2).\vec{r}_i}
e^{i(\vec{k}-\vec{q}/2).\vec{r}_j}\left(\bar{\chi}_{ij}+\delta\chi_{ij}(\tau)\right)
=\delta^D(\vec{q})n_{\vec{k}}+\rho_{\vec{k}}(\vec{q},\tau),
\end{equation}
where $D=2=$ dimension, $n_{\vec{k}}$ is the Fourier transform of $\bar{\chi}_{ij}$ and 
$\rho_{\vec{k}}(\vec{q},\tau)$ is the Fourier transform of $\delta\chi_{ij}(\tau)$. Notice that $(i,j)$'s are not restricted to nearest neighbor sites in the Fourier transforms.
It can be shown from the definition of $\chi_{ij}$ that $\rho_{\vec{k}}^*(\vec{q})=\rho_{\vec{k}}(-\vec{q})$. Similarly we define for the $\Delta_{ij}$ field,
\begin{equation}
\label{fourier2}
FT[\Delta_{ij}]=\sum_{i,j}e^{-i(\vec{k}+\vec{q}/2).\vec{r}_i}
e^{i(\vec{k}-\vec{q}/2).\vec{r}_j}\left(\bar{\Delta}_{ij}+\delta\Delta_{ij}(\tau)\right)=\delta^D(\vec{q})\Delta_{\vec{k}}+\Delta_{\vec{k}}(\vec{q},\tau),
\end{equation}
where $\Delta_{\vec{k}}$ is the Fourier transform of $\bar{\Delta}_{ij}$. It can be shown that $\Delta_{\vec{k}}(\vec{q})=\Delta_{-\vec{k}}(\vec{q})$. For the boson and constraint fields, we define
\begin{equation}
\label{fourier3}
FT[b(\lambda)_i]=\sum_ie^{i\vec{q}.\vec{r}_i}\left(\bar{b}(\bar{\lambda})+\delta{b}_i(\tau)(\delta\lambda_i(\tau))\right)=\delta^D(\vec{q})\sqrt{x}(\bar{\lambda})+b_{\vec{q}}(\tau)(\lambda_{-\vec{q}}(\tau)).
\end{equation}
\end{mathletters}

  Next we linear the equations of motion \ (\ref{eqmotion}). We consider small time-dependent fluctuations $\delta\chi_{ij}(\tau)$, $\delta\Delta_{ij}(\tau)$, etc. and keep to linear order time-dependent fluctuations in both sides of equations \ (\ref{eqmotion}).
Introducing Fourier transforms {\em via} equation \ (\ref{ft}), we obtain after some lengthy but straightforward algebra, six linearized equations of motion for the fluctuating fields $\rho_{\vec{k}}(\vec{q},\tau), \rho^+_{-\vec{k}}(-\vec{q},\tau), \Delta_{\vec{k}}(\vec{q},\tau), \Delta^+_{-\vec{k}}(-\vec{q},\tau), b_{-\vec{q}}(\tau)$ and $b^+_{\vec{q}}(\tau)$. The $\lambda_{\vec{q}}(\tau)$ field will be determined by the constraint equation. We shall simplify our notation by not displacing the explicit time-dependence of the fields entering the transport equations in the following. First we consider the equations of motion for the fermion fields.

 The equations of motion for the $\rho_{\vec{k}}(\vec{q})$ and $\Delta_{\vec{k}}(\vec{q})$ fields can be written in a matrix form. Introducing the vector notation
\[
\Psi_{\vec{k}}(\vec{q})= \pmatrix{\rho_{\vec{k}}(\vec{q}) \cr \rho_{-\vec{k}}^*(-\vec{q}) \cr \Delta_{\vec{k}}(\vec{q}) \cr \Delta^*_{-\vec{k}}(-\vec{q})},
\]
we obtain after Fourier transforming in time,
\begin{eqnarray}
\label{eqmf}
G_{0\vec{k}}^{-1}(\vec{q},i\omega)\Psi_{\vec{k}}(\vec{q}) & = & -{3J\over4}\sum_{\mu=\hat{x},\hat{y}}\cos(k_{\mu})\left(\Lambda_{\vec{k}}(\vec{q})\Delta_{\mu}(\vec{q})+\bar{\Lambda}_{\vec{k}}(\vec{q})\Delta^*_{\mu}(\vec{-q})\right)   \\  \nonumber
& & +H^s_{\vec{k}}(\vec{q})\left[\lambda_{\vec{q}}-\sum_{\mu}\cos(k_{\mu})\left(2t\cos({q_{\mu}\over2})\bar{b}(b_{-\vec{q}}+b^+_{\vec{q}})+{3J\over4}\rho^s_{\mu}(\vec{q})\right)\right]   \\  \nonumber
& & +H^a_{\vec{k}}(\vec{q})\sum_{\mu}\sin(k_{\mu})\left(2t\sin({q_{\mu}\over2})\bar{b}(b_{-\vec{q}}-b^+_{\vec{q}})+2it\bar{b}^2A^{ext}_{\mu}(\vec{q})-{3J\over4}\rho^a_{\mu}(\vec{q})\right),
\end{eqnarray}
where
\begin{eqnarray}
\label{g0}
G_{0\vec{k}}^{-1}(\vec{q},i\omega)=\pmatrix{i\omega-\xi_{\vec{k}+\vec{q}/2}+
\xi_{\vec{k}-\vec{q}/2} & 0 & {3J\over8}\bar{\Delta}_{\vec{k}+\vec{q}/2} & -{3J\over8}\bar{\Delta}_{\vec{k}-\vec{q}/2}  \cr 0 & i\omega-\xi_{\vec{k}-\vec{q}/2}+
\xi_{\vec{k}+\vec{q}/2} & {3J\over8}\bar{\Delta}_{\vec{k}-\vec{q}/2} & -{3J\over8}\bar{\Delta}_{\vec{k}+\vec{q}/2}  \cr {3J\over8}\bar{\Delta}_{\vec{k}+\vec{q}/2} & {3J\over8}\bar{\Delta}_{\vec{k}-\vec{q}/2} & i\omega+\xi_{\vec{k}+\vec{q}/2}+
\xi_{\vec{k}-\vec{q}/2} & 0  \cr -{3J\over8}\bar{\Delta}_{\vec{k}-\vec{q}/2} & -{3J\over8}\bar{\Delta}_{\vec{k}+\vec{q}/2} & 0 & i\omega-\xi_{\vec{k}+\vec{q}/2}-
\xi_{\vec{k}-\vec{q}/2}},\,\,\,
\end{eqnarray}

\[
\Lambda_{\vec{k}}(\vec{q})=\pmatrix{-\Delta_{\vec{k}+\vec{q}/2} \cr -\Delta_{\vec{k}-\vec{q}/2} \cr n_{\vec{k}+\vec{q}/2}+n_{\vec{k}-\vec{q}/2}-1 \cr 0} \, \, \, \, \,
\bar{\Lambda}_{\vec{k}}(\vec{q})=\pmatrix{\Delta_{\vec{k}-\vec{q}/2} \cr \Delta_{\vec{k}+
\vec{q}/2} \cr 0 \cr 1-n_{\vec{k}+\vec{q}/2}-n_{\vec{k}-\vec{q}/2}},\,\,
\]

and
\[
H^s_{\vec{k}}(\vec{q})= \pmatrix{n_{\vec{k}-\vec{q}/2}-n_{\vec{k}+\vec{q}/2} \cr n_{\vec{k}+\vec{q}/2}-n_{\vec{k}-\vec{q}/2} \cr -\Delta_{\vec{k}+\vec{q}/2}-\Delta_{\vec{k}-\vec{q}/2} \cr \Delta_{\vec{k}+\vec{q}/2}+\Delta_{\vec{k}-\vec{q}/2}} \, \, \, \, \,
H^a_{\vec{k}}(\vec{q})= \pmatrix{n_{\vec{k}-\vec{q}/2}-n_{\vec{k}+\vec{q}/2} \cr n_{\vec{k}-\vec{q}/2}-n_{\vec{k}+\vec{q}/2} \cr \Delta_{\vec{k}-\vec{q}/2}-\Delta_{\vec{k}+\vec{q}/2} \cr \Delta_{\vec{k}-\vec{q}/2}-\Delta_{\vec{k}+\vec{q}/2}},\,\,
\]
where $\rho^s_{\mu}(\vec{q})=\sum_{\vec{k}}\cos(k_{\mu})\rho_{\vec{k}}(\vec{q})$, $\rho^a_{\mu}(\vec{q})=\sum_{\vec{k}}\sin(k_{\mu})\rho_{\vec{k}}(\vec{q})$ and $\Delta_{\mu}(\vec{q})=\sum_{\vec{k}}\cos(k_{\mu})\Delta_{\vec{k}}(\vec{q})$. Writing $\Delta_{i,i+\hat{\mu}}=s_{\mu}|\Delta_{\mu}(i+{\hat{\mu}\over2})|e^{i\phi_{\mu}(i+{\hat{\mu}\over2})}$, where $s_{\hat{x}(\hat{y})}=+(-)1$, we obtain with Eq.\ (\ref{fourier2}) for small fluctuations $\Delta_{i,i+\hat{\mu}}\sim\bar{\Delta}_{\mu}+s_{\mu}\delta|\Delta_{\mu}|(i+{\hat{\mu}\over2})+i\bar{\Delta}_{\mu}\phi_{\mu}(i+{\hat{\mu}\over2})$, and
\begin{mathletters}
\label{pmeaning}
\begin{equation}
\label{pmdelta}
\Delta_{\mu}(\vec{q})=s_{\mu}\delta|\Delta_{\mu}|(\vec{q})+i\bar{\Delta}_{\mu}\phi_{\mu}(\vec{q}),
\end{equation}
showing that the real and imaginary parts of $\Delta_{\mu}(\vec{q})$ represent the amplitude and phase fluctuations of the superconducting order parameter, respectively. Similarly, writing $\chi_{i,i+\hat{\mu}}=|\chi_{\mu}|(i+{\hat{\mu}\over2})e^{ia_{\mu}(i+{\hat{\mu}\over2})}$, we obtain for small fluctuations,
\begin{equation}
\label{pmchi}
\rho^s_{\mu}(\vec{q})=\delta|\chi|_{\mu}(\vec{q}),  \,\,\, \rho^a_{\mu}(\vec{q})=i\bar{\chi}a_{\mu}(\vec{q})
\end{equation}
\end{mathletters}
showing that $\rho^{s(a)}_{\mu}(\vec{q})$ represents the amplitude (phase) fluctuations of $\chi$. Notice that $a_{\mu}(\vec{q})$ is identified as an internal gauge field in gauge theory\cite{le1}. It is also proportional to the mean-field fermion paramagnetic current, which is given by $j^p_{\mu}(\vec{q})=-i({3J\over4}+2t\bar{b}^2)\rho^a_{\mu}(\vec{q})$.

   The structure of the fermion matrix transport equation is quite similar to transport equations for usual Fermi liquid superconductors if only spin-singlet excitations are considered\cite{ysaul}. The absence of spin-triplet excitation in the matrix transport equation is a consequence of SBMFT which consider only decouplings in spin-singlet channel. $G_{0\vec{k}}^{-1}$ is the standard kinetic term for Fermi-liquid superconductors\cite{ysaul} and the coupling to $\Delta$ and $\rho^{s(a)}$ fields in the equations of motion can be understood easily from standard many-body theory for fermions. They are required in a conserving approximation in treating fermion-fermion interaction\cite{bk}. In equilibrium SBMFT, the Heisenberg interaction is decoupled in two channels resulting in the Cooper pairing field and "Fock" self-energy. A conserving approximation requires that corresponding vextex corrections must exist when computing the particle-hole and particle-particle propagators. They are precisely the $\Delta$ and $\rho^{s(a)}$ terms in the equations of motion. For ordinary Fermi liquid superconductors inclusion of superconductor order parameter fluctuations is essential to restore the correct longitudinal density response for superconductors\cite{schrieffer} and inclusion of vertex correction coming from self-energy is essential to obtain the correct Fermi liquid interactions\cite{bk}. Similar results are obtained in SBMFT. The main difference between SBMFT and ordinary Fermi liquid superconductor is the presence of boson components an constraints in the system so that in addition to the above vertex corrections, fermions in SBMFT also couple to the boson and constraint fields in the equations of motion.  

  The two linearized equations of motion for the boson fields are
\begin{mathletters}
\label{eqb}
\begin{eqnarray}
\label{eb1}
(-i\omega+\epsilon(\vec{q}))b_{\vec{q}} & = & -\bar{b}^2V(\vec{q})(b_{\vec{q}}+b^+_{-\vec{q}})+{t\bar{b}\over{V}}\sum_{\vec{k}'}\gamma(\vec{k}'-\vec{q}/2)\rho^*_{\vec{k}'}(\vec{q})  
\\  \nonumber
& & +2it\bar{b}\bar{\chi}\sum_{\mu=\hat{x},\hat{y}}\sin({q_{\mu}\over2})A^{ext}_{\mu}(\vec{q})-\bar{b}(\lambda_{-\vec{q}}+\Phi^{ext}(-\vec{q}))
\end{eqnarray}
and
\begin{eqnarray}
\label{eb2}
(-i\omega-\epsilon(\vec{q}))b^+_{-\vec{q}} & = & \bar{b}^2V(\vec{q})(b_{\vec{q}}+b^+_{-\vec{q}})-{t\bar{b}\over{V}}\sum_{\vec{k}'}\gamma(\vec{k}'+\vec{q}/2)\rho^*_{\vec{k}'}(\vec{q})
\\   \nonumber
& & -2it\bar{b}\bar{\chi}\sum_{\mu=\hat{x},\hat{y}}\sin({q_{\mu}\over2})A^{ext}_{\mu}(\vec{q})+\bar{b}(\lambda_{-\vec{q}}+\Phi^{ext}(-\vec{q})).
\end{eqnarray}
\end{mathletters}

 It is convenient to introduce the density-phase representation for bosons $b_i=n_ie^{i\theta_i}$. Using Eq. \ (\ref{eqb}) we obtain for small fluctuations, $\bar{b}(b_{-\vec{q}}+b^+_{\vec{q}})=n(-\vec{q})$ and $\bar{b}(b_{-\vec{q}}-b^+_{\vec{q}})=2i\bar{b}^2\theta(-\vec{q})$, 
\begin{mathletters}
\label{ebfinal}
\begin{eqnarray}
\label{ebd}
n(-\vec{q}) & = & {-2\bar{b}^2\epsilon(\vec{q})\over(i\omega)^2-E_h(\vec{q})^2)}\left(2t\sum_{\mu}\cos({q_{\mu}\over2})\rho^s_{\mu}(\vec{q})-(\lambda_{\vec{q}}+\Phi^{ext}(\vec{q}))\right)
\\  \nonumber
& & +{2\bar{b}^2(i\omega)\over(i\omega)^2-E_h(\vec{q})^2)}2t\sum_{\mu}\sin({q_{\mu}\over2})\left(\rho^a_{\mu}(\vec{q})+i\bar{\chi}A^{ext}_{\mu}(\vec{q})\right),
\end{eqnarray}
and
\begin{eqnarray}
\label{ebc}
2i\theta(-\vec{q}) & = & {-2(i\omega)\over(i\omega)^2-E_h(\vec{q})^2)}\left(2t\sum_{\mu}\cos({q_{\mu}\over2})\rho^s_{\mu}(\vec{q})-(\lambda_{\vec{q}}+\Phi^{ext}(\vec{q}))\right)
\\  \nonumber
& & +{2(\epsilon(\vec{q})+2xV(\vec{q}))\over(i\omega)^2-E_h(\vec{q})^2)}2t\sum_{\mu}\sin({q_{\mu}\over2})\left(\rho^a_{\mu}(\vec{q})+i\bar{\chi}A^{ext}_{\mu}(\vec{q})\right).
\end{eqnarray}
\end{mathletters}

  In the linear response regime, the constraint field $\lambda_{\vec{q}}$ should be choosen such that the density fluctuations of the fermion field is exactly balanced by the density fluctuations of the boson field, i.e. $\rho(\vec{q})=\sum_{\vec{k}}\rho_{\vec{k}}(\vec{q})=-n(-\vec{q})$. With eq.\ (\ref{ebd}), we obtain for the $\lambda$ field,
\begin{eqnarray}
\label{solc}
\lambda_{\vec{q}} & = & -\Phi^{ext}(\vec{q})+2t\sum_{\mu}\cos({q_{\mu}\over2})\rho^s_{\mu}(\vec{q})-{1\over\chi_h(\vec{q},i\omega)}\rho(\vec{q})  \\  \nonumber
& & -({i\omega\over\epsilon(\vec{q})})2t\sum_{\mu}\sin({q_{\mu}\over2})\left(\rho^a_{\mu}(\vec{q})+i\bar{\chi}A^{ext}_{\mu}(\vec{q})\right),
\end{eqnarray}
where $\chi_h(\vec{q},i\omega)={2\bar{b}^2\epsilon(\vec{q})\over(i\omega)^2-E_h(\vec{q})^2}$ is the mean-field density-density response function of the holes\cite{le1}. Putting eq. \ (\ref{solc}) in Eq. \ (\ref{ebc}), we obtain
\begin{equation}
\label{cont}
2i\bar{b}^2\epsilon(\vec{q})\theta(-\vec{q})=-i\omega\rho(\vec{q})-
4\bar{b}^2t\sum_{\mu}\sin({q_{\mu}\over2})\left(\rho^a_{\mu}(\vec{q})+i\bar{\chi}A^{ext}_{\mu}(\vec{q})\right),
\end{equation}
which is the continuity equation. To see that we rewrite Eq.(13) as
\begin{mathletters}
\label{current}
\begin{equation}
\label{con1}
{\partial\over\partial\tau}\rho(\vec{q})+\sum_{\mu}2i\sin({q_{\mu}\over2})J_{\mu}
(\vec{q})=0,
\end{equation}
where
\begin{equation}
\label{con2}
J_{\mu}(\vec{q})=2\bar{b}^2t\bar{\chi}\left(2\sin({-q_{\mu}\over2})\theta(-\vec{q})-a_{\mu}(\vec{q})-A^{ext}_{\mu}(\vec{q})\right),
\end{equation}
\end{mathletters}
is the physical current. The current expression (14b) can also be obtained directly from the current expression derived from $t$-term, where $J_{ij}\sim-t\sum_{\sigma}\left(e^{iA^{ext}_{ij}}c^+_{i\sigma}c_{j\sigma}b_ib^+_j-c.c.\right)$. In mean-field approximation, $J_{ij}\sim-t\sum_{\sigma}\left(\bar{b}^2(<c^+_{i\sigma}c_{j\sigma}>-c.c.)+\bar{\chi}\bar{b}^2(e^{i(\theta_i-\theta_j+A^{ext}_{ij})}-c.c.)\right)$, which for small fluctuations is precisely Eq. (14b) noting that $<c^+_{i\sigma}c_{j\sigma}>\sim\bar{\chi}e^{ia_{ij}}$. 

  Notice that using Eq.\ (\ref{cont}), we can also write
\[
\lambda_{\vec{q}}=-\Phi^{ext}(\vec{q})+2t\sum_{\mu}\cos({q_{\mu}\over2})\rho^s_{\mu}(\vec{q})+{E_h(\vec{q})^2\over2\bar{b}^2\epsilon(\vec{q})}+i(i\omega)\theta(-\vec{q}). \]
Together with the constraint equation $\rho(\vec{q})=-n(-\vec{q})$, we can eliminate the boson density and constraint fields from the fermion equations of motion. The coupling to boson phase $\theta(-\vec{q})$ can be eliminated through a gauge transformation, $c_{i\sigma}\rightarrow c_{i\sigma}e^{-i\theta_i}$ and $b_i^+\rightarrow{b}_i^+e^{i\theta_i}$ which absorb the boson phase on the fermions. It is straightforward to show that under this transformation, the fermion field $\Psi_{\vec{k}}(\vec{q})$ in the transport equation transform in the linear response regime as follows:
\begin{eqnarray}
\label{gtransform}
\rho_{\vec{k}}(\vec{q}) & \rightarrow & \rho_{\vec{k}}(\vec{q})+i(n_{\vec{k}-\vec{q}/2}-n_{\vec{k}+\vec{q}/2})\theta(-\vec{q}),  \\  \nonumber
\Delta_{\vec{k}}(\vec{q}) & \rightarrow & -i(\Delta_{\vec{k}-\vec{q}/2}+\Delta_{\vec{k}+\vec{q}/2})\theta(-\vec{q}),   \\ \nonumber
a_{\mu}(\vec{q}) & \rightarrow & a_{\mu}(\vec{q})-2\sin(-{q_{\mu}\over2})\theta(-\vec{q}),
  \\  \nonumber
\phi_{\mu}(\vec{q}) & \rightarrow & \phi_{\mu}(\vec{q})-2\cos({q_{\mu}\over2})\theta(-\vec{q}).
\end{eqnarray}

   Together with Eqs. \ (\ref{gtransform}) and \ (\ref{eqmf}), it can be shown that the gauge transformation removes all the $\theta(-\vec{q})$ fields in the equations of motion for the transformed fermion fields. Putting together we obtain,
\begin{eqnarray}
\label{eqtran}
G_{0\vec{k}}^{-1}(\vec{q},i\omega)\Psi_{\vec{k}}(\vec{q}) & = & -{3J\over4}\sum_{\mu}\cos(k_{\mu})\left(\Lambda_{\vec{k}}(\vec{q})\Delta_{\mu}(\vec{q})+\bar{\Lambda}_{\vec{k}}
(\vec{q})\Delta^*_{\mu}(\vec{-q})\right)   \\  \nonumber
& & +H^s_{\vec{k}}(\vec{q})\left[-\Phi^{ext}(-\vec{q})+2t\sum_{\mu}\cos({q_{\mu}\over2})\rho_{\mu}^s(\vec{q})+{E_h(\vec{q})^2\over2\bar{b}^2\epsilon(\vec{q})}\rho(\vec{q})\right.
\\  \nonumber
& & +\left.\sum_{\mu}\cos(k_{\mu})\left(2t\cos({q_{\mu}\over2})\rho(\vec{q})-{3J\over4}\rho^s_{\mu}(\vec{q})\right)\right]   \\  \nonumber
& & +H^a_{\vec{k}}(\vec{q})\sum_{\mu}\sin(k_{\mu})\left(2it\bar{b}^2A^{ext}_{\mu}(\vec{q})-{3J\over4}\rho^a_{\mu}(\vec{q})\right),
\end{eqnarray}
where all terms have the same meaning as before except they are now expressed in terms of the transformed fermion fields. Notice that $E_h(\vec{q})^2/(2\bar{b}^2\epsilon(\vec{q}))=\epsilon(\vec{q})/2\bar{b}^2+V(q)$.

  Equation \ (\ref{eqtran}) is the main result obtained in this paper. We note that after eliminating the bosons the transport equations for the fermions have the same form as a ($\vec{q}$-dependent) transport equations for ordinary Fermi liquid superconductors. The dynamics of the bosons is completely eliminated reflecting their 'slave" nature. Besides coupling of fermions to superconducting order parameter fluctuations, the fermions couple to each other through Landau molecular-field type interactions. We can extract effective $\vec{q}$-dependent Landau interactions by comparing Eq. \ (\ref{eqtran}) with Landau transport equations for superconductors\cite{legett}. We obtain,
\begin{equation}
\label{landaui}
f_{\vec{k}\vec{p}}(\vec{q})={1\over2}\left({\epsilon(\vec{q})\over2\bar{b}^2}+V(q)-{3J\over4}\sum_{\mu}\cos(k_{\mu}-p_{\mu})+2t\sum_{\mu}\cos({q_{\mu}\over2})(\cos(k_{\mu})+\cos(p_{\mu}))\right),
\end{equation}
which corresponds to a rather regular Fermi liquid. Notice that the density-density Landau interaction diverges in the $\bar{b}^2\rightarrow0$ limit, which we shall see indicates the vanishing of compressibility in the insulating state. The effects of strong correlation will be discussed in more detail in the next section.

  We note that with Eq (17) we can also introduce quasi-particle scattering into the transport equation through usual Fermi liquid phenomenology\cite{np,mahan}. The life-time of a quasi-particle with momentum $\vec{p}$ is given by the Boltzmann equation expression\cite{np},
\[
{1\over\tau_{\vec{p}}}\sim2\pi\sum_{q}\sum_{k}W(\vec{p},\vec{k};\vec{q})\delta(E_f(\vec{p})+E_f(\vec{k})-E_f(\vec{p}+\vec{q})-E_f(\vec{k}-\vec{q}))n_{\vec{k}}(1-n_{\vec{p}+\vec{q}})(1-n_{\vec{k}-\vec{q}}),  \]
where the scattering rate $W(\vec{p},\vec{k};\vec{q})$ can be identified as the absolute value square of the scattering amplitude $|A(\vec{p},\vec{k};\vec{q})|^2$ and can be computed with our Landau interactions \ (\ref{landaui})\cite{np}. We recall that our transport equation is derived only for spin-singlet channel, and spin-triplet landau interactions are missing.

  For ease of analysis it is convenient to separate the coupling of fermions to 'longitudinal' and 'transverse' fields explicitly. To achieve that we work in the Coulomb gauge such that $A^{ext}_{\mu}(\vec{q})$ is purely transverse and we write
\begin{mathletters}
\label{lt}
\begin{equation}
\label{lt1}
a_{\mu}(\vec{q})=a^t_{\mu}(\vec{q})+a^l_{\mu}(\vec{q}),
\end{equation}
where $\rho^a_{\mu}(\vec{q})=i\bar{\chi}a_{\mu}(\vec{q})$ and
\begin{eqnarray}
\label{lt2}
a^l_{\mu}(\vec{q}) & = & \sum_{\eta}{\sin({q_{\mu}\over2})\sin({q_{\eta}\over2})\over\sum_{\nu}[\sin({q_{\nu}\over2})]^2}a_{\eta}(\vec{q}),  \\ \nonumber
a^t_{\mu}(\vec{q}) & = & \sum_{\eta}\left(\delta_{\mu\eta}-{\sin({q_{\mu}\over2})\sin({q_{\eta}\over2})\over\sum_{\nu}[\sin({q_{\nu}\over2})]^2}\right)a_{\eta}(\vec{q}).
\end{eqnarray}
\end{mathletters}

  We then perform a gauge transformation to rewrite the coupling of the fermion fields to  $a_{\mu}^l(\vec{q})$ in terms of a coupling to scalar field $\lambda^a_{\vec{q}}$. Using the continuity equation and keeping in mind that the boson phases are already absorbed in the fermions in Eq.\ (\ref{eqtran}) we obtain $\sum_{\mu}2i\sin({q_{\mu}\over2})J_{\mu}(\vec{q})=-4i\bar{b}^2t\bar{\chi}\sum_{\mu}\sin({q_{\mu}\over2})a_{\mu}(\vec{q})=i\omega\rho(\vec{q})$. With the gauge transformation relation $2\sin({q_{\mu}\over2})\lambda^a_{\vec{q}}=(1-z)i(i\omega)a^l_{\mu}(\vec{q})$, we obtain $\lambda^a_{\vec{q}} =-(1-z){(i\omega)^2\over2\bar{b}^2\epsilon(\vec{q})}\rho(\vec{q})$, where 
\[
z={2t\bar{b}^2\over2t\bar{b}^2+{3J\over4}\bar{\chi}}. \]

  Putting this back into the Equation of motion \ (\ref{eqtran}), we obtain an matrix equation of motion for fermions where the longitudinal and transverse interactions are separated,
\begin{mathletters}
\label{eqff}
\begin{eqnarray}
\label{eqfinal}
G_{0\vec{k}}^{-1}(\vec{q},i\omega)\Psi_{\vec{k}}(\vec{q}) & = & -{3J\over4}\sum_{\mu}\cos(k_{\mu})\left(\Lambda_{\vec{k}}(\vec{q})\Delta_{\mu}(\vec{q})+\bar{\Lambda}_{\vec{k}}
(\vec{q})\Delta^*_{\mu}(\vec{-q})\right)   \\  \nonumber
& & +H^s_{\vec{k}}(\vec{q})\left[-\Phi^{ext}(-\vec{q})+2t\sum_{\mu}\cos({q_{\mu}\over2})\rho_{\mu}^s(\vec{q})-{1\over\chi^R_h(\vec{q},i\omega)}\rho(\vec{q})\right.
\\  \nonumber
& & +\left.\sum_{\mu}\cos(k_{\mu})\left(2t\cos({q_{\mu}\over2})\rho(\vec{q})-{3J\over4}\rho^s_{\mu}(\vec{q})\right)\right]   \\  \nonumber
& & +H^a_{\vec{k}}(\vec{q})\sum_{\mu}\sin(k_{\mu})\left(2it\bar{b}^2A^{ext}_{\mu}(\vec{q})-{3J\over4}i\bar{\chi}a^t_{\mu}(\vec{q})\right),
\end{eqnarray}
where 
\begin{equation}
\label{bfin}
\chi^R_h(\vec{q},\omega)={2\bar{b}^2\epsilon(\vec{q})\over(1-z)(i\omega)^2-E_h(\vec{q})^2},
\end{equation}
\end{mathletters}
is a renormalized density-density response function for the holes. 

\section{solutions to the transport equations}
  In this section we shall study the solutions of the transport equations in both the normal and superconducting states, concentrated in the small wavevector $\vec{q}$ limit and neglecting the quasi-particle life-time effects. We shall first consider the density and (transverse) current responses in the normal state and shall compare our results with corresponding results in U(1) gauge theory\cite{le1}. We shall then study the solutions of the transport equation in more details where quasi-particle properties will be examined. A similar study will then be carried out for the superconducting state. 

  The full equations of motion for fermions involving the complete Landau interactions
(17) are too complicated and cannot be solved exactly. To have a qualitative understanding of the solutions we approximate the Landau interactions by keeping only density-density and current-current interactions, i.e. we approximate
\begin{equation}
\label{alani}
f_{\vec{k}\vec{p}}(\vec{q})\sim{1\over2}\left(U^{eff}(\vec{q})-{3J\over4}\sum_{\mu}\sin(k_{\mu})\sin(p_{\mu})\right),
\end{equation}
where $U^{eff}(\vec{q})\sim{\epsilon(\vec{q})^2\over2\bar{b}^2}+V(\vec{q})+a(\vec{q})t-bJ$, $a$ and $b$ constants of order O(1) obtained from averaging the third and fourth terms in the Landau interactions (17) over the fermi surfaces. The effect of the $at-bJ$ term is to renormalize the charge-charge interaction $V(q)$ by adding a repulsive short-ranged potential of order $\sim t$. After making this approximation the transport equations become
\begin{eqnarray}
\label{eqsim}
G_{0\vec{k}}^{-1}(\vec{q},i\omega)\Psi_{\vec{k}}(\vec{q}) & = & -{3J\over4}\sum_{\mu}\cos(k_{\mu})\left(\Lambda_{\vec{k}}(\vec{q})\Delta_{\mu}(\vec{q})+\bar{\Lambda}_{\vec{k}}
(\vec{q})\Delta^*_{\mu}(\vec{-q})\right)   \\  \nonumber
& & +H^s_{\vec{k}}(\vec{q})\left(-\Phi^{ext}(-\vec{q})-{1\over\chi^{RR}_h(\vec{q},i\omega)}\rho(\vec{q})\right)  \\  \nonumber
& & +H^a_{\vec{k}}(\vec{q})\sum_{\mu}\sin(k_{\mu})\left(2it\bar{b}^2A^{ext}_{\mu}(\vec{q})-{3J\over4}i\bar{\chi}a^t_{\mu}(\vec{q})\right).
\end{eqnarray}
where $\chi^{RR}_h(\vec{q},i\omega)$ is given by Eq. (19b) with a renormalized charge-charge interaction $V^R(q)\sim V(q)+at-bJ$. We shall study this simplified matrix transport equation in the following. 

\subsection{solution in the normal state}
  In the normal state the fermion matrix equations of motion become a single equation 
\begin{eqnarray}
\label{emnormal}
(\omega-\xi_{\vec{k}+\vec{q}/2}+\xi_{\vec{k}-\vec{q}/2})\rho_{\vec{k}}(\vec{q}) & = & 
(n_{\vec{k}-\vec{q}/2}-n_{\vec{k}+\vec{q}/2})\times\left[-\Phi^{ext}(-\vec{q})-{1\over\chi^{RR}_h(\vec{q},i\omega)}\rho(\vec{q})\right.  \\  \nonumber
& & \left.+\sum_{\mu}\sin(k_{\mu})\left(2it\bar{b}^2A^{ext}_{\mu}(\vec{q})-{3J\over4}i\bar{\chi}a^t_{\mu}(\vec{q})\right)\right].
\end{eqnarray}

  In the small $\vec{q}$ limit the longitudinal and transverse responses are decoupled from each other and the density and current response functions can be solved easily. We obtain for the density-density response function,
\begin{equation}
\label{solnd}
\chi^d(\vec{q},\omega)={\chi_s^d(\vec{q},\omega)\chi^{RR}_h(\vec{q},\omega)\over
 \chi_s^d(\vec{q},\omega)+\chi^{RR}_h(\vec{q},\omega)}, 
\end{equation}
where $\chi^d(\vec{q},\omega)$ is the density-density response function of the system and
$\chi_s^d(\vec{q},\omega)$ is the corresponding free fermion response function, i.e., $\chi_s^d(\vec{q},\omega)=\sum_{\vec{k}}\chi_{0\vec{k}}(\vec{k},\omega)$, where
\[
\chi_{0\vec{k}}(\vec{k},i\omega)={n_{\vec{k}-\vec{q}/2}-n_{\vec{k}+\vec{q}/2}\over i\omega-\xi_{\vec{k}+\vec{q}/2}+\xi_{\vec{k}-\vec{q}/2}}.  \]

  To compute the current-current response function, we note that the physical current (14b) can be written as $J_{\mu}(\vec{q})=-zj_{\mu}^f=-z(j_{\mu}^p+j_{\mu}^d)$ after the bosons are eliminated, where $j_{\mu}^p=({3J\over4}+2t\bar{b}^2)\bar{\chi}a_{\mu}(\vec{q})$ and $j^d_{\mu}=(2t\bar{b}^2+{3J\over4}\bar{\chi})\bar{\chi}A^{ext}_{\mu}(\vec{q})$ are the fermion paramagnetic and diamagnetic currents, respectively. With this we obtain from solving equation \ (\ref{emnormal})
\begin{equation}
\label{solnj}
\chi^t(\vec{q},\omega)={\chi_s^t(\vec{q},\omega)\chi^t_h(\vec{q},\omega)\over
 (1-z)\chi_s^t(\vec{q},\omega)+\chi^t_h(\vec{q},\omega)}
\end{equation}
where $\chi^t(\vec{q},\omega)$ is transverse current-current response functions of the system, and $\chi_s^t(\vec{q},\omega)=\sum_{\vec{k}}[({3J\over4}+2t\bar{b}^2)\sin(k_{\mu})]^2\chi_{0\vec{k}}(\vec{q},\omega)$ is the corresponding free fermion response function. $\chi^t_h(\vec{q},\omega)=-2\bar{b}^2t\bar{\chi}$. We have written our results in a way which can be compared with gauge theory results easily\cite{le1,le2}. It is obvious that the transport equation produces essentially the same result as Ioffe-Larkin result for the density and current response functions, except the additional $z$ factors and the renormalization of density-density interaction $V^R(\vec{q})$. 

 We now examine the solutions of the transport equation in more detail. Following Fermi liquid theory, we consider solutions of the transports equation as eigen-excitations of the system\cite{np,shankar}. There are in general two kinds of solutions, dressed particle-hole excitations and collective modes\cite{np}. With our simplified Landau interaction, there exists only collective modes associated with density fluctuations in the system which can be determined from the poles of the density response functions. At small $\vec{q}$, the dispersion of the collective mode goes as $\omega_{\vec{q}}\sim\sqrt{2\bar{b}^2\epsilon(\vec{q})(V^R(\vec{q})+N(0)^{-1})/(1-z)}$, where $N(0)$ is the density of states on Fermi surface. Writing $V(q)\sim tv(q)$ where $v(q)$ is dimensionless, we obtain for small $x$ $\omega_{\vec{q}}$ goes as $t\sqrt{qx}$ for Coulomb interaction, and goes as $tq\sqrt{x}$ for short ranged interaction. Well defined collective mode exists at $|\vec{q}|<x(t/J)^2$ for long-ranged Coulomb interaction. Otherwise the collective mode is overdamped by the particle-hole continuum.

  We next consider solutions corresponding to dressed particle-hole pair excitations. In the normal state, we look for solutions of form\cite{np}
\[
\rho_{\vec{k}}(\vec{q})=\rho^{(0)}_{\vec{k}}(\vec{q})+\delta\rho_{\vec{k}}(\vec{q}), \]
where $\rho^{(0)}_{\vec{k}}(\vec{q})=\delta(\omega-\xi_{\vec{p}+\vec{q}/2}+\xi_{\vec{k}-\vec{q}/2})\delta^D(\vec{p}-\vec{k})$, corresponding to a bare particle-hole pair with relative momentum $\vec{p}$ and energy $\xi_{\vec{p}+\vec{q}/2}-\xi_{\vec{k}-\vec{q}/2}$. Putting this into the equation of motion \ (\ref{emnormal}) we obtain
\begin{equation}
\label{quasi}
\rho_{\vec{k}}(\vec{q})=-\rho^{(0)}_{\vec{k}}(\vec{q})+\chi_{o\vec{k}}(\vec{q},\omega) 
\left[\Phi^{ext}(-\vec{q})+{1\over\chi^{RR}_h(\vec{q},i\omega)}\rho(\vec{q})-\sum_{\mu}\sin(k_{\mu})\left(2it\bar{b}^2A^{ext}_{\mu}(\vec{q})-{3J\over4}i\bar{\chi}a^t_{\mu}(\vec{q})\right)\right].
\end{equation}

  The properties of the quasi-particles can be determined from solution of Eq. (25) if we identify $\rho_{\vec{k}}(\vec{r})$ as a local quasi-particle density as in Fermi-liquid theory. it is easy to solve Eq. (25) to obtain
\begin{equation}
\label{cquasi}
\rho(\vec{q})={\chi_h^{RR}(\vec{q},\xi_{\vec{p}+\vec{q}/2}-\xi_{\vec{k}-\vec{q}/2})\over
\chi_h^{RR}(\vec{q},\xi_{\vec{p}+\vec{q}/2}-\xi_{\vec{k}-\vec{q}/2})+\chi_s(\vec{q},\xi_{\vec{p}+\vec{q}/2}-\xi_{\vec{k}-\vec{q}/2})}\rho^{(0)}(\vec{q}),  
\end{equation}
where $\rho(\vec{q})=\sum_{\vec{k}}\rho_{\vec{k}}(\vec{q})$ is the total charge carried by the dressed particle-hole pair\cite{np} and $\rho^{(0)}(\vec{q})\sim1$ is the charge carried by the bare particle-hole pair. Overall speaking we find that the charge carried by the eigen-particle-hole pair (and the corresponding quasi-particle) is renormalized by a factor of order $\chi_h^{RR}\sim x$ from its bare value because of screening\cite{np}. For long-range Coulomb interaction the charge carried by the dressed particle-hole pair vanishes in the small $\vec{q}$ and $\omega=0$ limit. Similarly, we can also compute the (transverse) current carried by quasi-particle from Eq. (25). We obtain
\begin{equation}
\label{jquasi}
J^t(\vec{q})={\chi_h^t(\vec{q},\xi_{\vec{p}+\vec{q}/2}-\xi_{\vec{k}-\vec{q}/2})\over
\chi_h^t(\vec{q},\xi_{\vec{p}+\vec{q}/2}-\xi_{\vec{k}-\vec{q}/2})+(1-z)\chi_s^t(\vec{q},\xi_{\vec{p}+\vec{q}/2}-\xi_{\vec{k}-\vec{q}/2})}J^{(0)t}(\vec{q}),  
\end{equation}
where $J^{(0)t}(\vec{q})$ is the transverse current carried by the bare particle-hole pair. As in charge response, the overall current carried by the quasi-particle is renormalized by a factor of order $\chi^t_h\sim x$. However, in the small $\vec{q}$ and $\omega=0$ limit, the free fermion transverse current response function $\chi^t_s$ vanishes because of gauge invariance whereas $\chi^t_h$ remains finite $\sim x$, and the transverse current carried by quasi-particle is unrenormalized from its bare value.

  Our results can be understood readily from a Fermi liquid picture. With our simplified Landau interaction the long wavelength properties of the system are governed by the effective mass $m^*/m$ and two Landau parameter $F_0\sim V^R(0)N(0)$ (for short-ranged interaction) and $F_1=(z-1)$.  The effective mass correction is equal to ${m^*\over m}={t\over t\bar{b}^2+(3J/8)\bar{\chi}}$, which is of order $t/J$. Notice that the Fermi liquid identity $1+F_1={m^*\over m}$ which is rigorously satisfied for translationally invariant systems\cite{np,legett} is not satisfied here. Microscopically $F_1$ is coming from the vertex correction corresponding to the "Fock" decoupling of the Heisenberg interaction and vertex correction associated with slave-boson renormalization of bandwidth $t\rightarrow\bar{b}^2t$ is absent in SBMFT. 

  It is straightforward to see that our density-density response function is consistent with Fermi liquid theory. The static density-density response function is given by
\[
\chi^d(\vec{q},0)={\chi_s^d(\vec{q},0)\over1-U^{eff}(\vec{q})\chi_s^d(\vec{q},0)}, \]
in agreement with Fermi liquid theory\cite{np}. Notice that $\chi^d(\vec{q},0)$ goes to zero in the Mott insulator limit $x\rightarrow0, (lq)^2>>1$, where $l=1/\sqrt{x}$ is the average distance between holes, reflecting the vanishing of compressibility in the insulator state. Notice that external impurity potentials $V^I(\vec{q})$'s are screened by the factor $1-U^{eff}(\vec{q})\chi_s^d(\vec{q},0)$ correspondingly, implying that the effect of impurity scattering on quasi-particles is weakened by strong correlation and vanishes as the Mott insulator state is approached\cite{tjdlee}. 

  For the current response, the current expression (14b) implies that the physical current operator written in terms of the fermion current operator is given by $J_{\mu}(\vec{q})=(1+F_1)j^f_{\mu}(\vec{q})$, in agreement with Fermi liquid theory\cite{np,legett}. Our results for the current response can be obtained from Fermi liquid theory with this identification\cite{np,legett}. Notice that the "back-flow" current represented by $F_1$ exactly cancels the "bare" current in the Mott-insulator limit ($z\sim x\rightarrow0$), implying that the quasi-particles carry zero current in this limit, which seems to be consistent with an insulator phenomenology. 

\subsection{solution in the superconducting state}
   Next we analyse the fermion matrix transport equation in the superconducting state with the approximate Landau interactions (20).

  As in the normal state we study electromagnetic responses of the system in the small $\vec{q}$ limit where the density and transverse current responses are decoupled. First we consider the density response. With the simplified form of Landau interaction, the matrix transport equation becomes
\begin{eqnarray}
\label{eqfinal}
G_{0\vec{k}}^{-1}(\vec{q},i\omega)\Psi_{\vec{k}}(\vec{q}) & = & -{3J\over4}\sum_{\mu}\cos(k_{\mu})\left(\Lambda_{\vec{k}}(\vec{q})\Delta_{\mu}(\vec{q})+\bar{\Lambda}_{\vec{k}}
(\vec{q})\Delta^*_{\mu}(\vec{-q})\right)   \\  \nonumber
& & +H^s_{\vec{k}}(\vec{q})\left[-\Phi^{ext}(-\vec{q})-{1\over\chi^{RR}_h(\vec{q},i\omega)}\rho(\vec{q})\right],
\end{eqnarray}
which represents responses of a "pure" d-wave superconductor with no Landau interaction, to an effective external potential $-\Phi^{ext}(-\vec{q})-{1\over\chi^{RR}_h(\vec{q},i\omega)}\rho(\vec{q})$. Defining $\chi_{ds}^d(\vec{q},\omega)$ as the density-density response function of the pure d-wave superconductor, we find that the density-density response function of the system $\chi^d(\vec{q},\omega)$ is again give by equation (23), except that $\chi_s^d(\vec{q},\omega)$ is replaced by the corresponding response function $\chi_{ds}^d(\vec{q},\omega)$ of a d-wave superconductor. Similar conclusion is also reached for the transverse current response, with $\chi^t_s(\vec{q},\omega)$ replaced by the superconducting counter part $\chi^t_{ds}(\vec{q},\omega)$ in Eq. (24). These results are consistent with a generalized U(1) gauge theory for the superconducting state\cite{le1,le2}. 

   As in U(1) theory, the superfluid density deduced from $\chi^t(0,0)$ is given by $\rho_S\sim a'x-b'x^2T$ at small $x$ and low temperature $T$, where $a'$ and $b'$ are numerical factors of order O(1)\cite{le2,le3,mgil,mi}, and the $b'x^2T$ term is in disagreement with experimental results on the London penetration depth. The additional $(1-z)$ factor that appears in our transport equation solution does not modify this conclusion qualitatively. The $x$-dependence in the superfluid density originates from the Fermi liquid renormalization of the quasiparticle current $J_{\mu}(\vec{q})=(1+F_1)j^f_{\mu}(\vec{q})$ and is a fundamental property of SBMFT\cite{np,legett}. The disagreement with experiments suggests a fundamental problem associated with SBMFT for the high-$T_c$ cuprates\cite{mi}. 

  To study charge response, we neglect the small dissipative effects coming from nodal fermions and assume
\[
\chi_{ds}^d(\vec{q},\omega)={Kq^2\over\omega^2-Kuq^2},  \]
where $K,u\sim J$ and are the phase stiffness and inverse compressibility for the "pure" d-wave superconductor, respectively. We obtain from Eq. (23) that the $K$ and $U$ factors are renormalized by Landau interactions at small $\vec{q}$ to $K\rightarrow(2xK/(2x+(1-z){K\over t\bar{\chi}}))\sim xt$ and $u\rightarrow u +V^R(0)$ for short-ranged interaction $V^R(\vec{q})$, in agreement with U(1) theory\cite{dlee}. For Coulomb interaction the acoustic plasmon mode remains in superconducting state at small enough $\vec{q}$. Notice that the strong renormalization of compressibility and screening of impurity potential n the limit $(lq)^2>>1$ remains qualitatively the same in the superconducting state and may be the explanation why the nanometer-scale gap inhomogeneity revealed by STM experiments on BSCCO surface\cite{pan} does not seem to introduce strong quasi-particle scattering effect\cite{tjdlee}.

  The quasi-particle properties in the superconducting state can be studied from the transport equations as in the normal state. We look for solutions of form
$\Psi_{\vec{k}}(\vec{q})=\Psi_{\vec{k}}^{(0)}(\vec{q})+\delta\Psi_{\vec{k}}(\vec{q})$,
where $\Psi_{\vec{k}}^{(0)}(\vec{q})=\delta(\omega-E_f(\vec{p}-\vec{q}/2)-E_f(\vec{k}-\vec{q}/2))\delta^D(\vec{p}-\vec{k})\alpha_{\vec{p}}(\vec{q})$, where $E_f(\vec{k})$'s are mean-field quasi-particle energies and  
\[
\alpha_{\vec{p}}(\vec{q})=\pmatrix{ v_{\vec{p}-\vec{q}/2}u_{\vec{p}+\vec{q}} \cr u_{\vec{p}-\vec{q}/2}v_{\vec{p}+\vec{q}}  \cr -v_{\vec{p}-\vec{q}/2}v_{\vec{p}+\vec{q}} \cr u_{\vec{p}-\vec{q}/2}u_{\vec{p}+\vec{q}}}, \]
where $u_{\vec{k}}^2={1\over2}(1+{\xi_{\vec{k}}\over E_f(\vec{k})})$ and $v_{\vec{k}}^2={1\over2}(1-{\xi_{\vec{k}}\over E_f(\vec{k})})$ are superconductor coherence factors. The solution represents a dressed pair-excitation in the superconducting state. Proceeding as before we find that the charge and transverse current carried by the dressed particle-hole pair are given by Eqs. (23) and (24) as in the normal state, except that the normal state fermion response functions $\chi_s^d$ and $\chi_s^t$ are replaced by the corresponding functions in the superconducting state $\chi_{ds}^d$ and $\chi_{ds}^t$, respectively. In particular, we observe that because of vanishing of paramagnetic current, the transverse current carried by a quasi-particle is now renormalized by a factor $z\sim x$, which results in the $b'x^2T$ dependence of the superfluid density. The origin of the $z$ factor can be chased back to the current renormalization $J_{\mu}(\vec{q})=(1+F_1)j^f_{\mu}(\vec{q})$ as we have discussed.

  Notice that besides carrying charge and current, quasi-particles also carry with them disturbances in the mean-field order parameters $\chi_{ij}$ and $\Delta_{ij}$. It is found in SU(2) theory that low energy collective modes associated with fluctuations of the $\chi_{ij}$ and$\Delta_{ij}$ order parameters exist at momentum around $(\pi,\pi)$ and may enrich quasi-particle properties considerably\cite{ln}. These effects are not considered with our simplified form of Landau interaction (20).

\section{conclusion and comments}
  We now summarize our findings in this paper and discuss a few questions associated with our theory.

  By deriving complete transport equations we show in this paper how Gaussian fluctuations described by SBMFT of $t-J$ model can be understood in the language of Fermi liquid theory. An effective Fermi liquid Hamiltonian is derived which form the basis of our calculation of response function and quasi-particle properties. Our theory provides a more accurate description of Gaussian fluctuations in $t-J$ model compared with U(1) gauge theory and has the advantage that calculations beyond Gaussian theory (e.g. various transport coefficients) can be conveniently performed within the framework of Fermi liquid phenomenology. Unfortunately the main drawback of gauge theories associated with quasi-particle current renormalization is not remedied in our treatment. Notice that our formulation can be generalized to the high temperature phase where bose-condensation is absent in mean-field theory with a modified mean-field decoupling scheme where $<b_i>=0$ but $<b_i^+b_j>, <b_ib_j>\neq0$. It will be interesting to find out whether the high temperature phase can be identified as some kind of Fermi Liquid phase at high temperature in this approach. 

  A more fundamental question is to examine general features in our theory which are not specific to the $t-J$ model. To be specific we consider a lattice fermion model with hopping to both nearest and next-nearest neighbor sites and with (repulsive) density-density and spin-spin interactions different from the $t-J$ model. The only common feature of the model to $t-J$ model is the presence of strong-correlation handled by introducing slave-bosons. It is not difficult to see that the equations of motion for bosons will be essentially the same as the corresponding equations in $t-J$ model. except that the mean-field dispersion $\epsilon(\vec{q})$ and density-density interaction $V(q)$ are different. For repulsive interactions the behaviour of bosons will be very similar to bosons in $t-J$ model. The expression for $\lambda_{\vec{q}}$-field in linear response regime will also be very similar to those in $t-J$ model, since it is insensitive to the exact form of the spin-spin interaction. Therefore the main difference comes in the fermion part where a large variety of mean-field phases are possible, depending on details of the spin-spin interaction. After elimination of boson phase a generalized Fermi-liquid type description for fluctuations around the fermion mean-field state will be obtained for the lattice model, the precise forms of the Landau interactions will depend on the microscopic Hamiltonian and the mean-field fermion state.

  Notice however we expect that the Landau interactions will have some common features as in $t-J$ model. First of all, we expect that an density-density interaction term of form $f_{\vec{k}\vec{q}}^d(\vec{q})\sim E_h(\vec{q})^2/2\bar{b}^2\epsilon(\vec{q})$ will remain, since it comes from eliminating the bosons and constraint fields and does not depend on the fermion field. In fact, the presence of this term guarantees that the system becomes incompressible in the $x\rightarrow0$ limit. Secondly, we expect that current-current interaction term $F_1\sim ax-1$, where $a$ is a numerical constant of order O(1) will remain, as long as the fermion mean-field bandwidth remains finite in the limit $x\rightarrow0$. The existence of this factor in SMBFT can be inferred from mean-field decoupling of the hopping terms which always introduce a factor $\sim x$ in the coupling to external EM field. As a result quasi-particle carries vanishing current in the Mott Insulator limit $x\rightarrow0$. Notice that these are the two main features that dictates the density and current response of the SBMFT treatment in the small $\vec{q}$ limit. The temperature dependence of London penetration depth in high-$T_c$ cuprates suggests that current carried by quasi-particles on the Fermi surface does not vanish in the $x\rightarrow0$ limit. It would be interesting to examine whether this feature is unique to high-$T_c$ cuprates or whether there exists systems where quasi-particle current indeed vanished continuously when approaching the Mott-Insulator state.

  Finally we note that U(1) formalism is believed to be insufficient when describing properties associated slightly broken SU(2) symmetry. For simplicity we have concentrated at region $\vec{q}\sim(0,0)$ in this paper where SU(2) effect is not strong. To incorporate SU(2) fluctuations correctly a SU(2) version of transport equations is required. 

\acknowledgements
  The author acknowledges P.A. Lee and D.H. Lee for many helpful discussions.

\appendix
\section{Ioffe-Larkin formula from transports equation}

  To make direct contact with U(1) gauge theory\cite{le1} we consider the equations of motion before elimination of bosons. First we consider the current expression (14b). We note that the current expression can be interpreted as a boson (hole) current under the influence of effective gauge field $a_{\mu}(\vec{q})+A^{ext}_{\mu}(\vec{q})$. Alternatively we can rewrite the current expression as
\begin{equation}
\label{jf}
J_{\mu}(\vec{q})=-(2\bar{b}^2t+{3J\over4}\bar{\chi})\bar{\chi}\left(a_{\mu}(\vec{q})-(1-z)a_{\mu}(\vec{q})+z(A^{ext}_{\mu}(\vec{q})-a^b_{\mu}(\vec{q}))\right),
\end{equation}
where $a^b_{\mu}(\vec{q})=2\sin(-q_{\mu}/2)\theta(-\vec{q})$ is an effective gauge field coming from boson phase fluctuations. The current expression can be interpreted as (-ve) a fermion current under the influence of effective gauge field $(1-z)a_{\mu}(\vec{q})-z(A^{ext}_{\mu}(\vec{q})-a^b_{\mu}(\vec{q}))$. The existence of two interpretations for the physical current is a basic feature of U(1) gauge theory.

  We shall consider the normal state for simplicity. The superconducting state can be treated in a similar way. In this case $\Delta_{\vec{k}}=0$ and the fermion matrix transport equation is reduced to a single equation for the $\rho_{\vec{k}}(\vec{q})$ field. We further neglect direct coupling of the bosons and fermions to the $\rho^s_{\mu}(\vec{q})$ and $n(-\vec{q})$ fields in these equations and compute the density and current responses in both boson and fermion current interpretations. Using Eq.\ (\ref{ebfinal}) we obtain for the density and transverse current in "boson"-representation (in Coulomb gauge),
\begin{eqnarray}
\label{cboson}
n(-\vec{q}) & = & \chi_h(\vec{q},\omega)(\lambda_{\vec{q}}+\lambda^a_{\vec{q}}+\Phi^{ext}(\vec{q})),  \\  \nonumber
J^t_{\mu}(\vec{q})& = & \chi^t_h(\vec{q},\omega)(a^t_{\mu}(\vec{q})+A^{ext}_{\mu}(\vec{q})),
\end{eqnarray}
where $\lambda^a_{\vec{q}}$ comes from gauge transforming the longitudinal part of the $a_{\mu}(\vec{q})$ field. $a^t_{\mu}(\vec{q})$ is the transverse part of $a_{\mu}(\vec{q})$. Using Eq.\ (\ref{eqmf}), we obtain for the density and transverse currents in "fermion"-representation,
\begin{eqnarray}
\label{cfermion}
\rho(\vec{q}) & = & \chi_s^d(\vec{q},i\omega)(\lambda_{\vec{q}}+(1-z)\lambda^a_{\vec{q}})
\\   \nonumber
J_{\mu}(\vec{q}) & = & -\chi^t_s(\vec{q},i\omega)((1-z)a^t_{\mu}(\vec{q})+zA^{ext}_{\mu}(\vec{q})),
\end{eqnarray}
In the $z\rightarrow0$ limit, we can neglect the $O(z)$ terms in equation \ (\ref{cfermion}) and the two equations can be solved self-consistently easily. We obtain the well known Ioffe-larkin expression for the density and current response functions,
\[
\chi^{d}(\vec{q},\omega)={\chi_h(\vec{q},\omega)\chi_s^d(\vec{q},\omega)\over\chi_h(\vec{q},\omega)+\chi_s^d(\vec{q},\omega)}, \]
and 
\[
\chi^{t}(\vec{q},\omega)={\chi_h^t(\vec{q},\omega)\chi_s^t(\vec{q},\omega)\over\chi_h^t(\vec{q},\omega)+\chi_s^t(\vec{q},\omega)}.  \]
A parallel analysis can also be done for the superconducting state.

\end{document}